\documentclass[traditabstract]{aa}
\usepackage{amsmath}
\usepackage{txfonts}
\usepackage{graphicx}
\usepackage{natbib}
\usepackage{amsfonts}
\usepackage{amssymb}
\usepackage{hyperref}
\usepackage{subfigure}
\usepackage[flushleft]{threeparttable}


\bibpunct{(}{)}{;}{a}{}{,} 

\renewcommand{\hat}[1]{\oldhat{\mathbf{#1}}}

\begin{document}

\title{Surface rotation of \emph{Kepler} red giant stars\footnote{ables n and n' are only available in electronic form
at the CDS via anonymous ftp to cdsarc.u-strasbg.fr (130.79.128.5) or via http://cdsweb.u-strasbg.fr/cgi-bin/qcat?J/A+A}}
\author{T. Ceillier \inst{1,2} \and
J. Tayar \inst{3} \and
S. Mathur \inst{4} \and
D.~Salabert \inst{1,2} \and
R.~A. Garc\'\i a\inst{1,2} \and  
D.~Stello\inst{5,6}\and
M.~H. Pinsonneault \inst{3} \and
J. van Saders \inst{7,8} \and
P.~G.~Beck \inst{1,2} \and
S. Bloemen \inst{9}}

\institute{IRFU, CEA, Universit\'e Paris-Saclay, F-91191 Gif-sur-Yvette, France
\and Universit\'e Paris Diderot, AIM, Sorbonne Paris Cit\'e, CEA, CNRS, F-91191 Gif-sur-Yvette, France
\and Department of Astronomy, Ohio State University, 140 W 18th Ave, OH 43210, USA
\and Center for Extrasolar Planetary Systems, Space Science Institute, 4750 Walnut Street, Suite 205, Boulder, CO 80301, USA
\and Sydney Institute for Astronomy (SIfA), School of Physics, University of Sydney, NSW 2006, Australia
\and School of Physics, University of New South Wales, NSW 2052, Australia
\and Carnegie-Princeton Fellow, Carnegie Observatories, 813 Santa Barbara Street, Pasadena, California, 91101 USA
\and Department of Astrophysical Sciences, Princeton University, Princeton, NJ 08544, USA
\and Department of Astrophysics, IMAPP, Radboud University Nijmegen, PO Box 9010, NL-6500 GL Nijmegen, The Netherlands
}

\abstract{   {\textit{Kepler} allows the measurement of starspot variability in a large sample of field red giants for the first time. With a new method that combines autocorrelation and wavelet decomposition, we measure 361 rotation periods from the full set of 17,377 oscillating red giants in our sample. This represents 2.08\% of the stars, consistent with the fraction of spectroscopically detected rapidly rotating giants in the field. The remaining stars do not show enough variability to allow us to measure a reliable surface rotation period. Because the stars with detected rotation periods have measured oscillations, we can infer their global properties, e.g. mass and radius, and quantitatively evaluate the predictions of standard stellar evolution models as a function of mass. Consistent with results for cluster giants when we consider only the 4881 intermediate-mass stars, M$>$2.0 M$_{\odot}$ from our full red giant sample, we do not find the enhanced rates of rapid rotation  expected from angular momentum conservation. We therefore suggest that either enhanced angular momentum loss or radial differential rotation must be occurring in these stars. Finally, when we examine the 575 low-mass (M$<$1.1 M$_{\odot}$) red clump stars in our sample, which were expected to exhibit slow (non-detectable) rotation, 15\% of them actually have detectable rotation. This suggests a high rate of interactions and stellar mergers on the red giant branch.} }


\maketitle

\section{Introduction}
   {Isolated low-mass red giant stars are expected to be inactive and slowly rotating. They lose angular momentum through a magnetized wind on the main sequence, and are then further slowed by the increase in the star's moment of inertia as its envelope expands on the red giant branch \citep{weber1967, schatzman1962, 1972ApJ...171..565S, 2003ApJ...586..464B, MamajekHillenbrand2008}. However, in spectroscopic samples of field stars, it appears that about 2\% exhibit rapid rotation  \citep{1993ApJ...403..708F, Massarotti2008, 2011ApJ...732...39C,1996A&A...314..499D}. Two explanations are commonly put forth for such stars. The first is that not all stars are isolated: 44\%\  of low-mass stars form in binaries \citep{Raghavan2010}, and more than one-quarter of these stars are expected to interact on the giant branch \citep{2011ApJ...732...39C}. An even larger fraction of stars are thought to have substellar companions, and these too can interact to produce a rapidly rotating star \citep{2016A&A...593A.128P, 2009ApJ...700..832C}. Given that many interesting classes of stars arise from binary interactions and mergers, including low-mass white dwarfs, cataclysmic variables, and Type Ia supernova, better empirical constraints on the rate of binary interactions are interesting for a wide range of applications. Stars resulting from a merger are easiest to identify in clusters \citep[e.g.][]{2016ApJ...832L..13L, 2004ApJ...604L.109P}, but recent work has also identified field giants with unusual chemistry for their age as likely merger products \citep{2015MNRAS.451.2230M}, although such stars are relatively rare. Since the orbital angular momentum of a binary  system can be transformed into spin angular momentum during the tidal interaction and merger of the two bodies, identifying rapidly rotating red giants is a way to quantify the merger rate on the giant branch. Current population synthesis models suggest that between 1 and 2\%\ of red giants should be rapidly rotating on the giant branch due to interactions \citep{2011ApJ...732...39C}.} 

   {The second explanation for rapid rotation is that not all stars are low mass. Stars above the Kraft break on the main sequence ($\sim$6250 K, $\sim$1.3M$_{\odot}$) do not have substantial convective envelopes. They are therefore not expected to spin down substantially on the main sequence \citep{1978GApFD...9..241D}, and observations indicate that they are indeed still rotating rapidly (velocities up to 300 km/s) at the end of the main sequence \citep{Zorec2012}. Assuming solar-like angular momentum loss on the giant branch, we expect about half of these stars to still be fast enough to be detected during the core helium burning phase at rotation periods of tens of days (velocities above 10 km/s), although this theoretical prediction contrasts with recent results from open clusters \citep{2016ApJ...818...25C}.}

   {When we combine the expectations from interactions and massive stars we predict that significantly more than the measured two percent of stars should be rapidly rotating, which suggests that there could be a problem with the simple picture presented above. One of the most likely explanations is that the standard assumptions of solar-like spin-down rates are incorrect for giants, and that there are many moderately rotating (3-10 km/s) giants, but fewer rapidly rotating ($>$10 km/s) giants than predicted. This would have implications for our understanding of the mechanism and timescale of angular momentum transport  \citep[see e.g.][]{2013A&A...555A..54C, 2013ApJ...775L...1T,2014ApJ...788...93C}, mass and angular momentum loss \citep{Reimers1975}, and stellar magnetism \citep{2015Sci...350..423F,2016Natur.529..364S}. To determine whether this is the cause of the discrepancy would require measurements of the full distribution of rotation rates of all intermediate mass stars, including those rotating slowly.}

   {The other way to explain the discrepancy between the predicted and observed rates of rapidly rotating giants is that there are incorrect assumptions used when computing the merger rates of binary systems. While the fraction of stars in binaries is supposed to be well constrained, the rate of interactions is also sensitive to the distribution of mass ratios and binary separations, which are not well known \citep{DucheneKraus2013}.  In order to determine whether the merger rate assumptions are at fault, we would need a sample of stars of known mass because all low-mass (M$<$1.3 M$_{\odot}$) giants not undergoing any interaction should be rotating extremely slowly (periods of hundreds of days, velocities less than 1 km/s).}

   {Clearly, in order to test  these two explanations for the low fraction of rapidly rotating giants in the field, we need a large, homogeneous sample of single stars of known mass whose full rotation distribution, down to very low speeds, can be characterized. Such a sample would be difficult to obtain spectroscopically because measuring moderate and slow rotational broadening is difficult. It requires high-resolution high signal-to-noise spectra and a precise model of the turbulent broadening, which can be several kilometers per second in red giants. Additionally, while spectroscopic measurements of mass do exists \citep{2016MNRAS.456.3655M, 2016ApJ...823..114N}, they are indirect and tend to have large uncertainties (up to 0.2 M$_{\odot}$).}
   
   We therefore focus on photometric measurements of our red giant sample. Using photometry to measure rotation is still challenging because these stars tend to have periods from tens to hundreds of days, and are expected to have low-amplitude modulations due to magnetic variability. While a large sample of such measurements would be challenging to obtain from the ground, it is well matched to the observations already obtained by the \emph{Kepler} satellite, which has more than 1400 days of observations of $\sim 17000$ field giants at millimagnitude precision. The very good quality of these photometric measurements allows the determination of the stellar surface rotation through the periodic variations of brightness of an active star induced by the magnetic spots crossing over the visible disk \citep[e.g.][]{2009A&A...506...33M, 2009A&A...506...41G, 2010A&A...518A..53M, 2012A&A...548L...1D,2012A&A...543A.146F,2014A&A...564A..50L,2016ApJ...823...16B}. Various methods using this principle have been developed and have led to the detection of surface rotation for a large number of stars in the \emph{Kepler} field \citep[e.g.][]{2013MNRAS.432.1203M, 2013ApJ...775L..11M,2014ApJS..211...24M,2013A&A...557L..10N,2014A&A...572A..34G,2016MNRAS.456..119C}. However, for observational reasons, most of these surveys have focused on dwarfs with rotation periods typically below 100 days.


   The \emph{Kepler} photometric data also allows the measurement of masses of field red giants through the technique of asteroseismology. These stars undergo stochastically excited solar-like oscillations, and the frequency of maximum power of these oscillations ($\nu_{\rm max}$) and the spacing between modes of the same spherical degree and consecutive radial order ($\Delta \nu$) can be combined to infer the mass and surface gravity of each star using scaling relations \citep{1991ApJ...368..599B,1995A&A...293...87K}. Having the mass of each star will help us distinguish between low-mass stars that are rotating rapidly due to a recent interaction and stars rotating rapidly because they were born with a mass above the Kraft break.  

In the present work, we study the surface rotation of the most complete sample of red giants
observed by the \emph{Kepler} satellite. In Sect.~\ref{Sec:sample}, we describe our stellar sample and the preparation of the light curves while in Sect.~\ref{Sec:surf}, we detail how the extraction of surface rotation is carried out. Our results and their implications are discussed in Sect.~\ref{Sec:disc} and our conclusions are summarized in Sect.~\ref{Sec:Conclu}.

\section{Sample selection and data correction}
\label{Sec:sample}
\begin{figure}[!htb]
\begin{center}
\includegraphics[width=9cm]{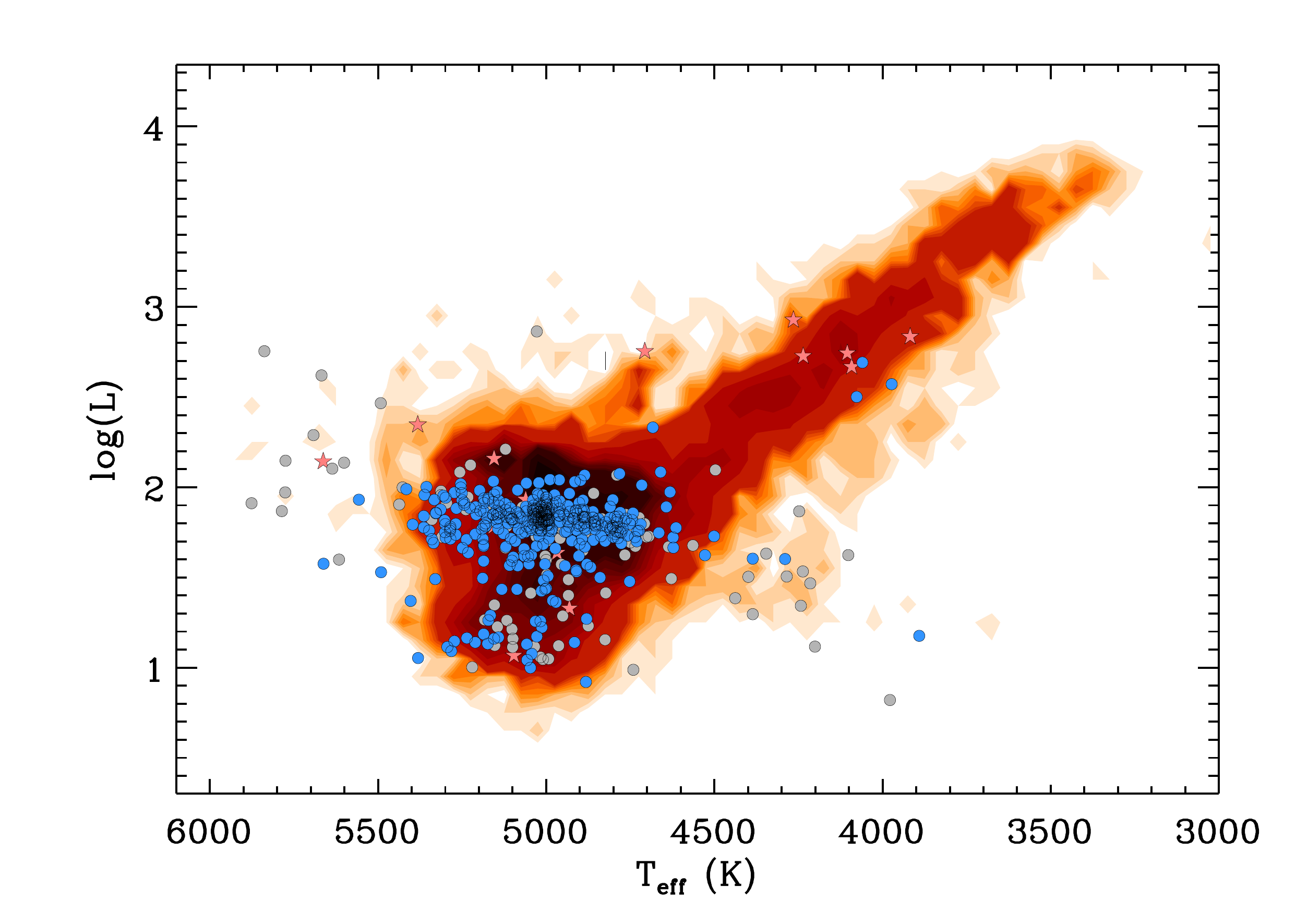}
\caption{Hertzsprung-Russell diagram of the full sample of  red giants. The density map
  corresponds to the whole sample (17,377 stars). The grey dots are
  stars for which a rotational modulation is detected but that are
  discarded as probable pollution (151 stars). The magenta stars show the positions of the 19 stars removed according to the $T_\text{crit}$ criterion. The blue dots represent stars for which a reliable rotation period has been derived (361 stars). See Section~\ref{Sec:surf} for details.}
\label{Fig:1}
\end{center}
\end{figure}

 As only a few red giants are supposed to exhibit light curve modulations due to star spots,  for this work we use the largest sample of identified red giants observed by the \emph{Kepler} satellite to date. It is composed of 17,377 pulsating stars including those already known from previous works 
\citep[e.g.][]{2010ApJ...723.1607H,2011MNRAS.414.2594H,2012A&A...537A..30M,2013ApJ...765L..41S,2016ApJ...827...50M}. The global seismic parameters $\nu_\text{max}$ and $\Delta\nu$ are computed in a homogeneous way using the A2Z seismic pipeline \citep{2010A&A...511A..46M} and are used to infer the stellar masses using the seismic scaling relations 
($\Delta\nu \propto \rho$, $\nu_\text{max} \propto$ g/T$_{\rm{eff}}^{0.5}$) \citep{1995A&A...293...87K}. Our sample contains, in particular, 4881 intermediate-mass stars with M$>$2.0 M$_{\odot}$ and 575 low-mass clump stars with M$<$1.1 M$_{\odot}$. The distribution in the Hertzsprung-Russell (HR) diagram  of the full set of 17,377 red giants can be seen in Fig.~\ref{Fig:1}.

For each star, the longest available observations recorded by the \emph{Kepler} mission are used, i.e. from Q0 to Q17
spanning  1470 days starting May 2, 2009, and ending May 11, 2013. Because we are interested in the
surface rotation periods that are low-frequency modulations -- typically
with periods longer than a day -- only long cadence data with a sampling
rate of 29.4244  min (Nyquist frequency of 283.45 $\mu$Hz) are
used. None of the two available NASA data products, Simple Aperture
Photometry (SAP) or Pre-search Data Conditioning multi-scale Maximum A
Posterior methods (PDC-msMAP) \citep{ThompsonRel21} can be directly 
used. However, SAP light curves have not been corrected for many
instrumental perturbations and the data of each quarter is not
normalized, while PDC-msMAP light curves are high-pass filtered with an attenuation starting at three-day periods that removes essentially all of the signal above 20 days 
\citep[e.g.][]{{ThompsonRel21}}. Although the latest \emph{Kepler}
data releases re-inject part of the identified stellar long-period
signal back into the light curves, it is not guaranteed that this is
done for all the quarters of a star or for all the stars in our sample
\citep[][]{2013ASPC..479..129G}. Therefore, we extract our own
aperture photometry from the pixel-data files following a
simple automatic algorithm. It starts by determining a reference value
for the amount of flux in a pixel as the 99.9th percentile of the flux
in the pixel during a full quarter (avoiding outliers). Then the
original mask is extended by moving away from the centre of the PSF in
all directions, and includes pixels as long as their reference value
is above a given threshold,  and on the condition that the reference
value drops while moving away from the centre. If a pixel has a flux
below the threshold the algorithm stops adding pixels in this
direction. If the flux starts to increase, which is a sign of the
presence of another star, the algorithm also stops adding pixels at
this point in this direction (for further details see Mathur et al., in prep.). Once the photometry of all the
quarters is extracted, we use the KADACS pipeline \citep[\emph{Kepler}
Asteroseismic Data Analysis and Calibration
Software,][]{2011MNRAS.414L...6G} to correct for outliers, jumps, and
drifts and to properly concatenate the independent
quarters. These data are then high-pass filtered using a triangular
smoothing function with three different cut-off periods at 20, 55, and
80 days, producing three different light curves. The first two  filters
are done by quarter, while the last  is applied to the full
series. To avoid border effects in the quarters when short cut-off
frequencies are selected we extend the light curve by assuming
symmetry with respect to each of the two ending points before applying
the filter. For the rest of the paper we only discuss the results from the 55 and 80-day filters.

Finally, the \emph{Kepler} data suffers from regular interruptions in
the data acquisition due to instrumental operations that produce a
regular window function which introduces high-frequency harmonics in
the power spectrum when it couples with high-amplitude low-frequency
modulations such as the rotation-induced modulations that we are trying to study
\citep[for more details see][]{2014A&A...568A..10G}. To minimize this
effect, all gaps shorter than 20 days are interpolated using
inpainting techniques \citep{2015A&A...574A..18P}. 
Because  these corrections are sometimes  not perfect, we remove from
the light curves the quarters that show an anomalously high variance
compared to that of their neighbours. To this end, we calculate for each star the variance of every
quarter and divide the resulting array by its median. Then, the
difference in this ratio between each quarter and its two neighbours
is computed. If the mean of these two differences is greater than a
threshold -- empirically set to 0.9 -- we remove the quarter from
the light curve. Finally, we rebin the light curve by a factor of 4 to
speed up the analysis. This does not affect the range of periods in which we are interested. 

\section{Studying the surface rotation}
\label{Sec:surf}

\subsection{Wavelets and ACF analyses}

\begin{figure*}[!htb]
\begin{center}
\includegraphics[width=18cm]{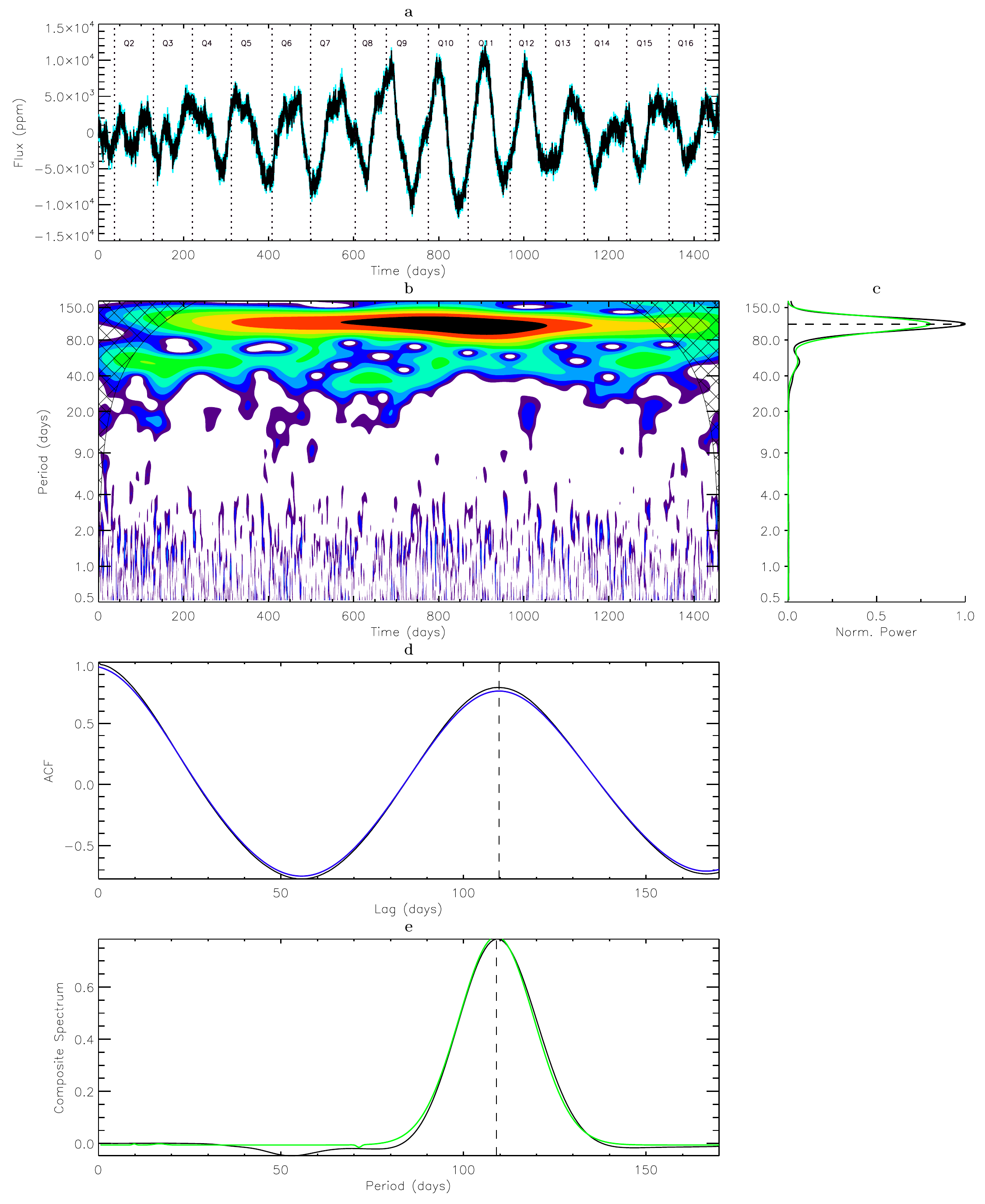}
\caption{Analysis of the light curve of KIC\,2436732, filtered at 80 days. Panel~a: Original light curve (light blue) and rebinned light curve (black). Panel~b: Wavelet decomposition (WPS) of the rebinned light curve. Panel~c: GWPS (black) and Gaussian fit (green). Panel~d: ACF of the rebinned light curve (black) and smoothed version of this ACF (blue). Panel~e: Composite spectrum of the rebinned light curve (black) and Gaussian fit (green). For each method, a dashed line indicates the returned period.}
\label{Fig:KIC2436732}
\end{center}
\end{figure*}

The methodology we apply here is similar to that used in
\citet{2014A&A...572A..34G} adapted to study red giant stars. 
One of the differences is that only one type of data -- KADACS corrected data -- is used, with two
different high-pass filters with the cuts at 55 and 80 days. The first step
of our methodology is computing a time-period analysis based on a
wavelet decomposition of the rebinned
light curve to obtain the wavelet power spectrum (WPS). The WPS can be used to see whether a modulation is due to a glitch or is present during the whole data set. We then project this WPS on the period axis to form the global wavelets power
spectrum (GWPS) which is similar to a Fourier spectrum but with a
reduced resolution. The advantage of the GWPS is that it increases the
power of the fundamental period of a signal
and thus avoids mistaking an overtone for the true
periodicity of the signal \citep[][]{2013A&A...549A..12M}. The GWPS is then described and
 minimized via the least-squares method using multiple Gaussian functions. The central
period of the Gaussian function corresponding to the highest peak is
then taken as the rotation period $P_{rot,GWPS}$. The half-width at
half-maximum of this function is taken as the uncertainty on this
value. In the case of red giants, the solar-like oscillations can have
periods of the order of a day and can be mistaken for rotation. To
prevent this, we use the measured power excess and we exclude the range
$[\nu_{max}-5\Delta\nu\ ;\ \nu_{max}+5\Delta\nu]$ from the search for a
rotation period. We have verified with stars of different $\nu_{max}$ and a wide range of magnitudes that this range is the minimum interval we need to remove to avoid any pollution of the rotation period measurements by the oscillation modes. This range is directly interpolated in the GWPS
prior to the fitting with Gaussian functions. Examples of the WPS
and the GWPS can be seen in Fig.~\ref{Fig:KIC2436732}.

The second step of the methodology is calculating the autocorrelation
function (ACF) of the light curve, following
\citet{2013ApJ...775L..11M}. The ACF is smoothed according to the
most significant period present in its Lomb-Scargle
periodogram. The smoothing is performed using a Gaussian function
that is  a tenth of the selected period in width. However, if this period is smaller than the one
corresponding to the frequency $\nu_{max}-5\Delta\nu$, that one is
taken instead. The significant peaks in the
smoothed ACF are then identified. The highest peak is taken as
the rotation period $P_{rot,ACF}$. The presence of a regular pattern
due to the presence of several active regions on the star at different
longitudes, as noted by \citet{2013ApJ...775L..11M}, is also
checked. If no significant peak is
identified, the star is considered inactive (or observed with a very
low inclination angle). An example of the ACF can be
seen in Fig.~\ref{Fig:KIC2436732}.

\subsection{Composite spectrum}

The third method is a combination of the two previous ones.  We create a new function, called the Composite Spectrum (CS), that is obtained by multiplying the ACF and the GWPS together \citep{2016MNRAS.456..119C}. This is done to boost the height of the peaks present in both curves and to decrease the height of the peaks present in only one of the two. As the ACF and the GWPS are not sensitive to the same problems in the original light curve, this allows us to  more easily identify periods intrinsic to the star.

To do so, we first fit the smoothed ACF with an exponentially
decreasing function of the form
\begin{equation}
  f_{exp}(P)=(1-A_0)\ exp(-P/A_1))+A_0 \;\text{,} 
\end{equation}
where $P$ is the period and $A_0$ and $A_1$ are the two parameters 
to fit. We note that $f_{exp}(0)=1$ and
$f_{exp}(\infty)=0$. This fit is then subtracted from the smoothed
ACF. The normalized ACF is then rebinned into the periods of the GWPS to allow the proper multiplication of the two quantities.

The CS is then obtained by multiplying the normalized
ACF by the normalized GWPS. This allows the composite spectra of all stars to have comparable
amplitude. The CS has approximately the same
resolution as the GWPS and is defined on the same periods. It
allows a direct reading of the relevant periods present in the
original light curve. As is true for the GWPS, the CS is fitted with multiple
Gaussian functions. The central period of the function corresponding
to the highest peak is taken as $P_{rot,CS}$ and the half width at
half maximum of this function is taken as the uncertainty on this
value. An example of the CS can be seen in Fig.~\ref{Fig:KIC2436732}.

This methodology has been tested on a large sample of composite light curves in a hare-and-hounds exercise and has been found to be one of the best compromises between completeness and accuracy among all the available methodologies \citep{2015MNRAS.450.3211A}. In this work, we apply it to the two high-pass filtered light curves with cut-off periods at 55 and 80 days, resulting in six rotation periods for each star: $P_{rot,GWPS}$, $P_{rot,ACF}$, and $P_{rot,CS}$ for each filter.

\subsection{Automatic selection of probable detections}

In the previous work detailed in \citet{2014A&A...572A..34G}, the
relatively small size of the sample allowed us to perform a visual
check of the results for each star. For the present work, the huge
size of the sample prevents us from using the same very time
consuming method. Moreover, only a small number of red giant stars
are expected to show light curve modulations due to spots. We
therefore need an automated way to extract from our initial sample a
subsample containing only the potentially rotating red giants.

To do so, we introduce three values, $H_{ACF}$, $H_{CS}$, and
$G_{ACF}$. The first is   $H_{ACF}$, which is the height of the peak in the
smoothed ACF that corresponds to the rotation period $P_{rot,ACF}$, as defined in
\citet{2013ApJ...775L..11M}. It is the mean of the difference between
the value of the smoothed ACF at the top of the peak and the values of
the smoothed ACF at the two neighbouring minima. The second, $H_{CS}$, is defined the same way but
for the CS. Finally, $G_{ACF}$ is the global maximum of the smoothed
ACF after it first crossed 0, for the range of periods considered. 
The reasons why we consider the first two values are obvious; however,
the use of $G_{ACF}$ needs to be explained. Its role is mostly to compensate for the way $H_{ACF}$ is computed, which does not give a precise idea of the degree of correlation between the light curve and
itself with a lag of $P_{rot, ACF}$ as the maximum of the peak can be
positive and small or even negative; the $H_{ACF}$ can still be high
if the neighbouring minima are very low. After testing several methods
of balancing these effects, $G_{ACF}$ is the most efficient and easiest to
compute.

\begin{figure}
\begin{center}
\includegraphics[width=9cm]{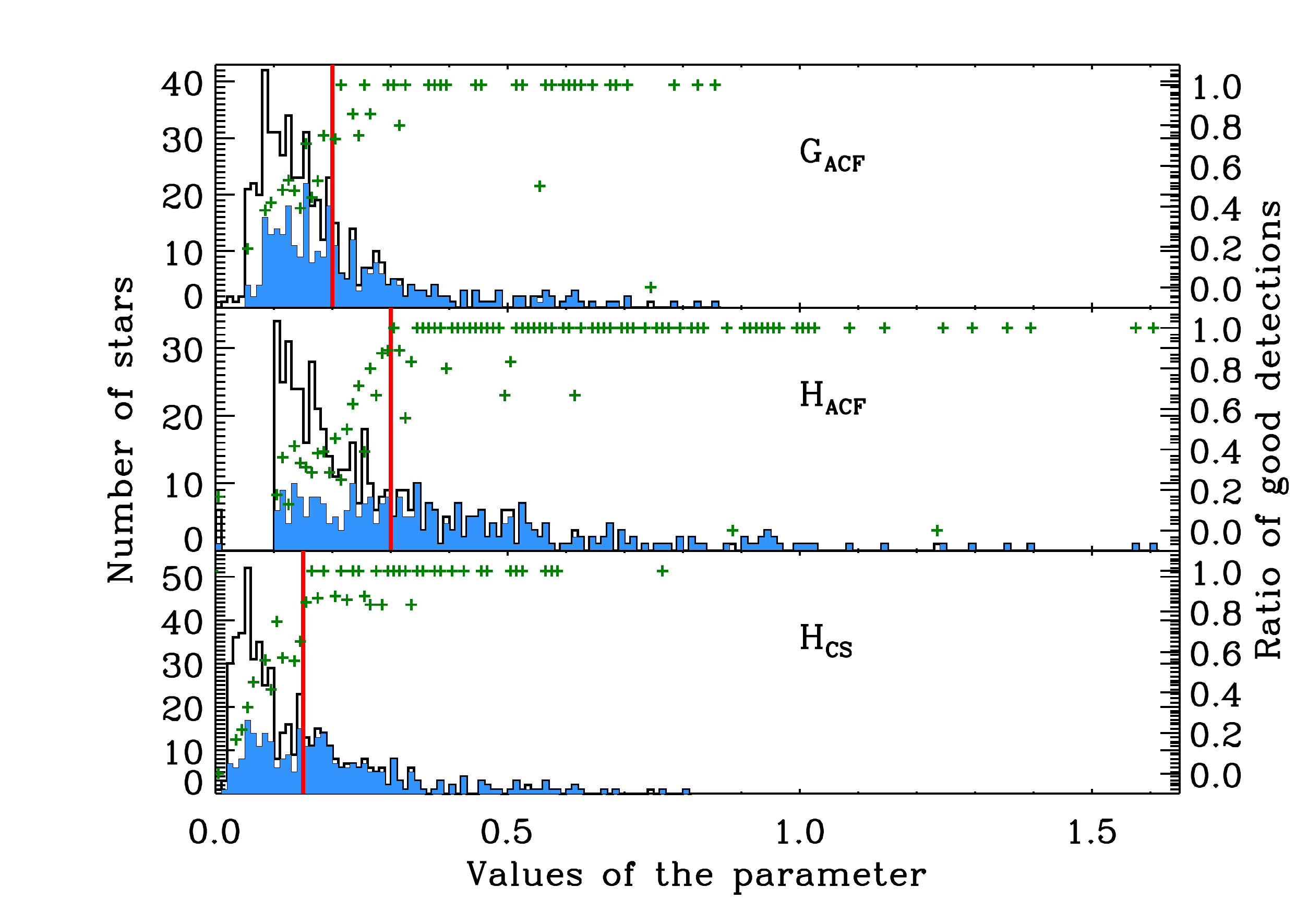}
\caption{Histograms of $G_{ACF}$, $H_{ACF}$, and $H_{CS}$ for the stars
studied by \citet{2014A&A...572A..34G}. Black lines: whole
sample (540 stars). Blue: stars with detected surface rotation (310 stars). The green crosses
represent the ratio of good detections for each bin and the red lines
mark the threshold value used for each parameter.}
\label{Fig:Y}
\end{center}
\end{figure}

To validate the use of these three values and decide which threshold we should use for each of them, we calculated $H_{ACF}$, $H_{CS}$, and
$G_{ACF}$ for the KADACS corrected light curves of the 540 solar-like stars in
\citet{2014A&A...572A..34G}. Figure~\ref{Fig:Y} shows the histograms of
these three values  for the whole sample of 540 stars and for the
310 stars with detected surface rotation. We find that if we select all the stars
with $H_{ACF}\ge0.3$, $H_{CS}\ge0.15$, and $G_{ACF}\ge0.2$, we recover
124 of the 310 stars showing rotation (40\%) and only 6 of the 230
stars showing no rotation (2.6\%). The key point here is that the
subsample thus obtained contains only 4.6\% ($6/130$) stars without
detected rotation. This selection method is then able to
isolate a sample of stars with a high probability of detectable surface
rotation.

Applying this selection method to our sample of 17,377 red giants, we isolate 925 stars for which the criteria $H_{ACF}\ge0.3$, $H_{CS}\ge0.15$, and $G_{ACF}\ge0.2$, are fulfilled for at least one of the two high-pass filtered light curves  used (cut-off periods at 55 or 80 days). From now on, we  only deal with this reduced sample. The remaining stars will be
considered to have a low probability of detectable surface rotation.

\subsection{Visual check of the subsample results}

As the \emph{Kepler} light curves can be sensitive to instrumental effects, we then visually check the light curve, GWPS, ACF, and CS for each of the 925 stars and for both filters. If the period detected by one of these analyses is also visible in the light curve as stellar signal, this period is kept as the rotation period of the star. In contrast, when the signatures in the different methods are not clear enough or come from instrumental effects, no period is returned.

When it appears that the period detected seems to be a harmonic of the
real rotation period, we apply a longer filter to the data and
re-do the rotation period extraction process. This can happen when
the first peak in the ACF is at a period below 100 days, but the second
and higher ones are at a period greater than 100 days. It can be due
to spots or active regions appearing on opposite sides of the star and
producing this characteristic pattern \citep[see][]{2013MNRAS.432.1203M}.

After this phase of visual inspection, only 531 stars out of the 925
of our subsample are kept. The light curves of all these stars
demonstrate clear rotational modulations. For these 531 red giants,
the final rotation period is taken from the fit of
the corresponding peak in the GWPS. The returned value $P_{rot}$ is
the period of the maximum of the Gaussian function fitted while the
uncertainty $\delta P_{rot}$ is the half width at half maximum of this
function. When possible, these values are taken from the GWPS of the 80-day filtered light curve. Otherwise, the GWPS of the 55-day filtered light curve is used.

\begin{figure*}[!htb]
\begin{center}
\includegraphics[trim=1.5cm .8cm 1.5cm 0.1cm, width=0.8\textwidth]{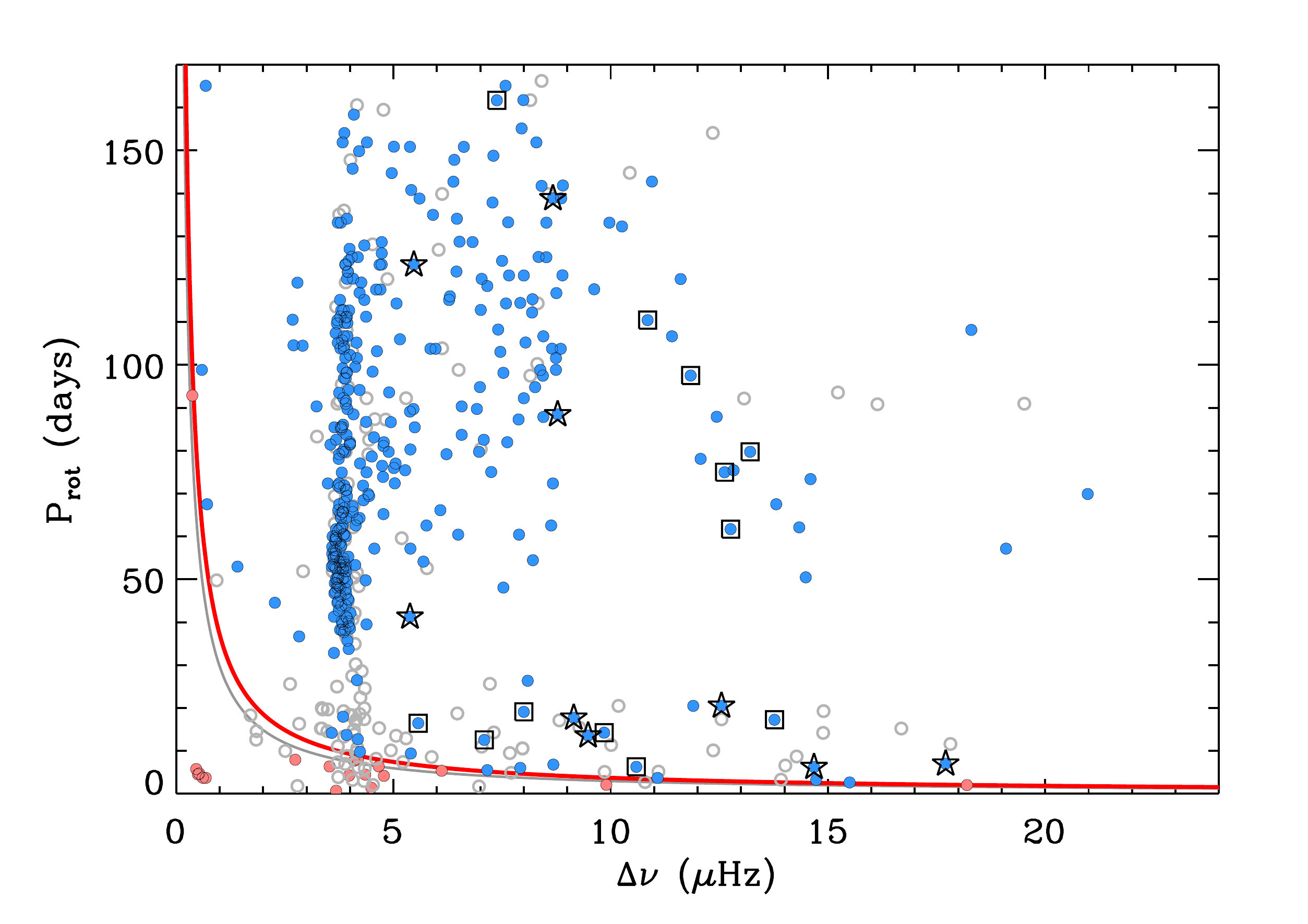}
\caption{  Distribution of stars showing rotational modulation in the $P_\text{rot}$-$\Delta\nu$ space. Open grey dots correspond to the 151 stars discarded according to the crowding criterion. Filled magenta dots are the 19 stars removed according to the $T_\text{crit}$ criterion. Blue dots represent stars for which a reliable rotation period has been derived (361 stars). The grey line indicates the critical period $T_\text{crit}$ and the red line marks a rotational velocity of 80\% of the critical value. The open square symbols indicate RGB stars with depressed dipolar modes following \citet{2016Natur.529..364S}, while star symbols indicate stars with normal dipolar modes.}
\label{Fig:crit}
\end{center}
\end{figure*}

\subsection{Discarding probable pollution}

In some cases, it is possible that the rotational modulation detected
is not produced by the observed red giant. This can happen for two
reasons: the red giant is in fact part
of a multiple system and the modulation is produced by an active
companion star or there is another active star that is close to the
red giant on the sky and whose light is contaminating the red giant's
light curve, i.e. a chance alignment \citep{2017arXiv170500621C}. In the first scenario, it is difficult to detect the presence of a companion without a detailed study of the star,
including spectroscopic observations to check for multiple spectral
lines or radial velocity variations. 

In the second scenario, which is more likely to be the dominant one as MS stars are likely to be very faint relative to giants,  it is possible to estimate the
probability that the mask used to compute the red giant's light cruve
contains signal from other stars. The \emph{Kepler} data
products contain a parameter called {crowding}\footnote{The crowding values used here were the ones provided by the MAST at https://\url{archive.stsci.edu/kepler/} on April 2015.}, ranging from 0 to
1, that corresponds to the fraction of the flux from the target star. In other words, the closer to 1 this parameter is, the
less polluted the light curve is. We thus discard all the red giants
among the 531 for which the {crowding} parameter is lower than
0.98, eliminating 151. Interestingly, the proportion of
low-{crowding} stars is very high among the red giants with
rotation periods of less than 30 days, which shows that most of these
detections are due to pollution of the light curves. In contrast, the
proportion of low-{crowding} stars is very low for red giants
with rotation periods above 100 days, which tends to validate these
detections.

\subsection{Comparison with breakup rotation periods}

Finally,  we compare the rotation rate of stars with the critical period $T_\text{crit}$ under which a star would
be torn apart by the centrifugal force. This period can be calculated
as 
\begin{equation}
  T_\text{crit}=\sqrt{\frac{27\pi^2R^3}{2GM}} \; \text{,}
\end{equation}
where $R$ and $M$ are the radius and the mass of the star and $G$ is
the gravitational constant. Using seismic scaling laws
\citep[see][]{1995A&A...293...87K}, this expression can be simplified as
\begin{equation}
  T_\text{crit}=\sqrt{\frac{27\pi^2{R_\odot}^3}{2GM_\odot}} \left(
    \frac{\Delta\nu}{\Delta\nu_\odot} \right)^{-1 }\; \text{,}
\end{equation}
where $R_\odot$, $M_\odot$, and $\Delta\nu_\odot$ are respectively the radius, the mass, and the large separation of the Sun,  and $\Delta\nu$ is the large separation of the star (see grey line in Fig. ~\ref{Fig:crit}).

For each of the 531~stars, we calculate this critical period. If the
measured period $P_\text{rot}$ is lower than 1.25~$T_\text{crit}$ --
corresponding to a star rotating at 80\% of its critical velocity (red line in Fig. ~\ref{Fig:crit}) --
we consider the measurement as doubtful. The modulation in the light curve could be due either to pollution of another star or to the presence of a companion in a binary system (causing modulation by the binary orbit and not by the rotation period of the giant star). 

The parameters $P_\text{rot}$ and $T_\text{crit}$ are illustrated in
Fig.~\ref{Fig:crit} as a function of $\Delta \nu$, which represents the mean density of the stars and provides some information on the evolutionary stage of the stars (red clump stars are located between 3 and 4.5 $\mu$Hz). We clearly see that many of the stars with a rotation period smaller than the critical period were already discarded by the pollution criterion. Among the 380 stars with a crowding above 0.98 (i.e. above the pollution criterion), we find that nineteen stars are rotating faster than the $T_\text{crit}$ criterion (magenta dots in Fig.~\ref{Fig:crit}).


In some other cases, it is also possible that the derived value of
$\Delta\nu$ is incorrect, leading to an incorrect value of
$T_\text{crit}$. However, we can see in Figs.~\ref{Fig:crit} and ~\ref{Fig:hist} that some generally smaller red giants with a rotation periods below thirty days are not discarded by
this verification.

\subsection{Comparison with known Kepler binaries}

 We have cross-checked our sample of 531 stars showing rotational modulation with the Villanova binary catalog \citep{2016AJ....151...68K} containing all the known binaries in the \emph{Kepler} main mission. We find that only one star is a binary: KIC 5990753, with an orbital period of 7.2 days\footnote{http://keplerebs.villanova.edu/overview/?k=5990753}.

In addition, we have also cross-checked our data set with the list of \emph{Kepler} compact binary systems around red giants studied by \cite{2017arXiv170500621C}. From a total of 168 stars analysed in that paper, thre are only ten stars  in common with our 531 red giants. Six of them are possible true compact binaries, but only one, KIC~12003253, is retained in our final list of 361 `confirmed rotation' red giants. The other five were flagged as  1 or 2. The reported peak associated with the orbital modulation of the secondary star in this system is 2 days and our rotation period for the red giant is 54 days. Therefore, we think the value reported in our paper is the true rotation period of the red giant.  The other four stars in common were reported as possible chance alignments or pollution in Colman's paper. Three of these four stars are also flagged in our analysis (flag 1). The last one, KIC~7604896, is polluted by a signal with a period of 0.16 days as indicated in \citet{2017arXiv170500621C}, while our rotation period is 88.46 days. Therefore, we also think the value we reported is the rotation of the red giant.

\subsection{Link between surface rotation and mode suppression}
It has been proposed that the reduction in power of non-radial modes observed in red giants
\citep{2012A&A...537A..30M,2017A&A...598A..62M,2014A&A...563A..84G,2016PASA...33...11S} arises
from magnetic suppression \citep{2015Sci...350..423F,2017MNRAS.467.3212L}. The magnetic fields
are thought to have been formed in the progenitor stars during the main sequence
phase by a dynamo established due to the convection and rotation in the
core regions \citep{2016Natur.529..364S}. If we assume that the surface rotation in the red giant phase correlates with the core rotation during the main sequence, we might expect that rapidly rotating red giants might have had faster main sequence core rotation rates and thus be more likely to have suppressed non-radial modes. To study this potential link, we cross-matched the RGB stars investigated by \citet{2016Natur.529..364S} for mode suppression (i.e. stars with $\Delta \nu$ greater than $5\,\mu$Hz to avoid including clump stars) with our RGB stars with $\Delta \nu$ above the same threshold and with reliable rotation measurements (131 stars). We found 12
stars that show suppression in our sample (blue filled dots surrounded by an
open blue square in Fig.~\ref{Fig:crit}), and 9 stars with normal
oscillation mode power (blue filled dots surrounded by an open star symbol
in Fig.~\ref{Fig:crit}). 
Of the 11 fast rotators ($P_\mathrm{rot}\,<\,30\,$days) in that sample, 6 show suppressed modes and 5 show a normal oscillation pattern. For comparison, of the 10 slow rotators in the sample ($P_\mathrm{rot}\,>\,30\,$days), there are 6 with depressed dipole modes, and 4 normal oscillators. 
We therefore see no obvious correlation between the two phenomena (rapid surface rotation and suppressed dipole modes) in our sample. Additional analysis should be done when a larger sample is available in order to draw stronger conclusions.

\section{Results and discussion}
\label{Sec:disc}

\begin{figure}
\begin{center}
\includegraphics[width=9cm]{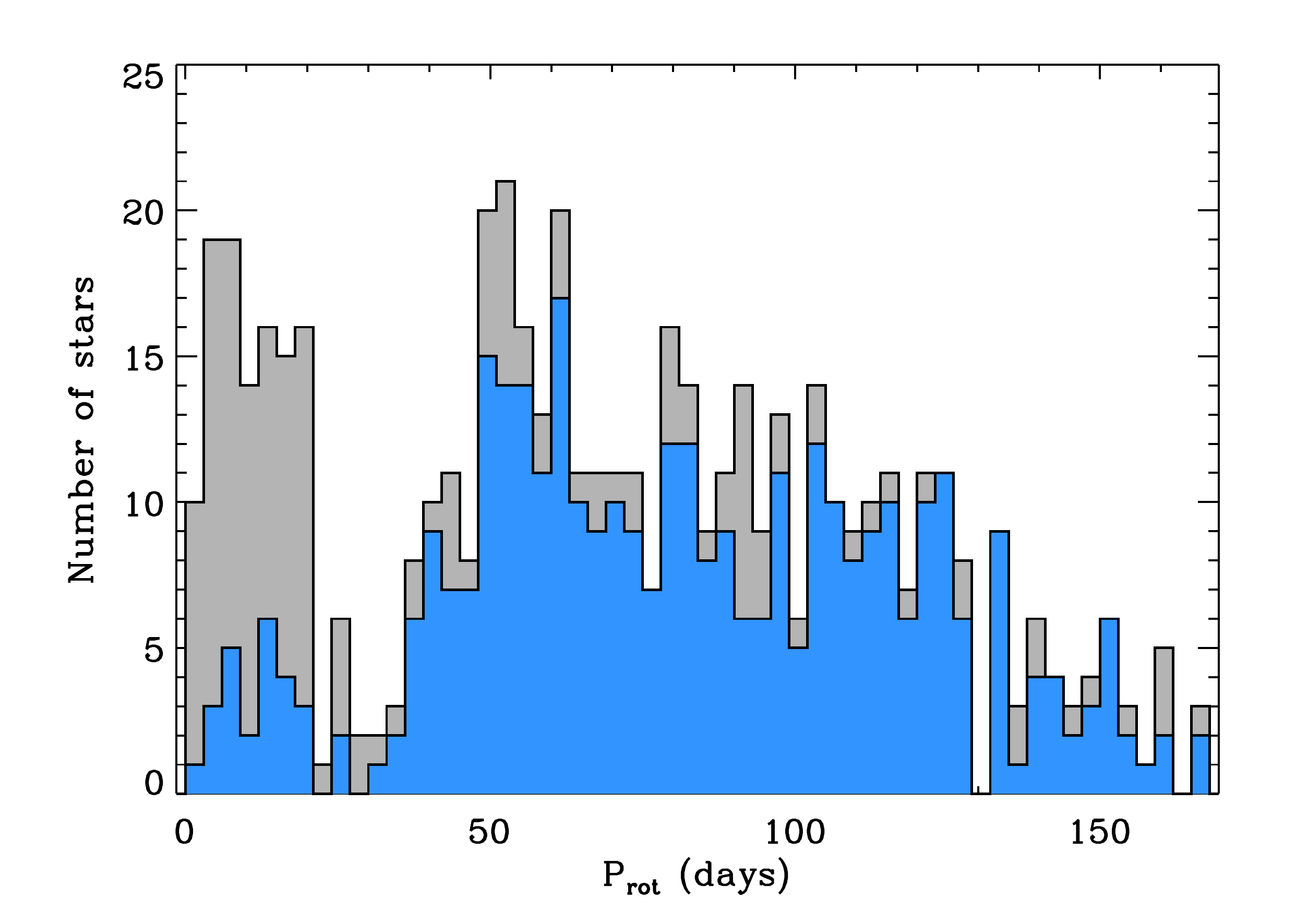}
\caption{Rotation periods $P_\text{rot}$ for all the
  stars showing rotational modulation (grey, 531 stars) and for stars
  for which a reliable rotation period has been derived (blue, 361 stars).}
\label{Fig:hist}
\end{center}
\end{figure}

We initially detected 531 stars showing signatures of rotational
modulation. {From this set we remove  151 stars  due to the crowding factor. We also discard 19 additional stars whose rotation rates are close to or faster than the critical break-up period. Thus,} we keep a sample of 361 stars with confirmed surface rotation periods. These
results are summarized in Table~\ref{Tab:Sample} and the distribution of
the derived $P_\text{rot}$ can be seen in Fig~\ref{Fig:hist}. It is clear that while short periods are more likely to be detectable, they are also more likely to be a signal from a companion or contaminant. In contrast, longer periods are more likely to be from the red giant, but we suspect that the fall-off at longer periods is a selection effect, as long-period, low-amplitude signals are the most difficult to detect. In these tables, we also give the values of the global parameters of the p modes ($\Delta \nu$ and $\nu_{\rm max}$) with the mass and surface gravity computed from the seismic scaling relations.

Now that the sample of stars is defined we can compare our results with spectroscopic measurements
and study the distribution of the active red giants that we detected
to better understand the underlying scenarios explaining this high
activity and rotation.

\begin{table*}
  \caption{\label{Tab:Sample} Stars with validated rotation periods.}
  \begin{threeparttable}
  \centering
  \begin{tabular}{c*{6}{r@{$\ \pm\ $}l}cc*{2}{r@{$\ \pm\ $}l}}
KIC & \multicolumn{2}{c}{$\nu_\text{max}$ [$\mu$Hz]} &
\multicolumn{2}{c}{$\Delta\nu$ [$\mu$Hz]} &
\multicolumn{2}{c}{T$_{\rm{eff}}$ [K]} & 
\multicolumn{2}{c}{$\log g$ [dex]} & 
\multicolumn{2}{c}{M [M$_\odot$]} & 
\multicolumn{2}{c}{$P_\text{rot}$ [days]} & 
crowding & $T_\text{crit}$ [days] &
\multicolumn{2}{c}{$v sini$ [km/s]} \\
\hline
1161618&     34.25&      1.88&      4.09&      0.14& 4855&  161& 1.29& 0.16&
 2.45&  0.03&158.34&   7.90 &  0.990&  7.26&   0.00&   0.00 \\
1162746&     27.58&      1.81&      3.64&      0.13& 5141&  308& 1.17& 0.18&
 2.37&  0.04& 46.74&   4.72 &  1.000&  8.16&   0.00&   0.00 \\
1871631&     29.29&      2.15&      3.71&      0.11& 4823&  108& 1.18& 0.17&
 2.38&  0.03& 44.54&   3.08 &  0.990&  8.01&   0.00&   0.00 \\
2018667&     30.73&      1.80&      3.83&      0.14& 4773&  123& 1.19& 0.15&
 2.40&  0.03& 65.67&   3.52 &  0.990&  7.75&   0.00&   0.00 \\
2019396&     37.98&      1.89&      4.16&      0.13& 4790&  100& 1.62& 0.18&
 2.49&  0.02& 63.86&   8.70 &  0.990&  7.14&   0.00&   0.00 \\
2156178&     29.08&      1.87&      3.93&      1.91& 5099&  225& 1.00& 0.98&
 2.39&  0.03& 40.98&   3.18 &  1.000&  7.56&   0.00&   0.00 \\
2305930&     28.27&      1.78&      3.97&      0.14& 4924&  173& 0.84& 0.11&
 2.37&  0.03& 33.75&   2.52 &  0.990&  7.48&  13.09&   0.88 \\
2436732&     30.34&      1.76&      3.70&      0.11& 4719&  147& 1.29& 0.16&
 2.39&  0.03&109.63&   5.17 &  0.980&  8.03&   0.00&   0.00 \\
2447529&    106.95&      4.87&      8.38&      0.18& 5186&  247& 2.47& 0.26&
 2.96&  0.03& 98.83&   5.49 &  0.990&  3.54&   0.00&   0.00 \\
2716214&     30.26&      1.95&      3.91&      0.12& 5120&  279& 1.16& 0.17&
 2.41&  0.03& 41.31&   2.07 &  0.990&  7.60&   0.00&   0.00 \\
2845408&     74.76&      3.95&      6.38&      0.12& 5058&  185& 2.42& 0.26&
 2.80&  0.03&142.71&  14.66 &  0.990&  4.66&   0.00&   0.00 \\
2845610&     90.10&      5.20&      6.97&      0.14& 5095&  151& 3.00& 0.34&
 2.88&  0.03& 79.72&   8.51 &  1.000&  4.26&   0.00&   0.00 \\
2854994&    133.16&     10.78&     10.26&      0.21& 5003&  117& 2.01& 0.30&
 3.05&  0.04&132.27&   6.54 &  1.000&  2.89&   0.00&   0.00 \\
2856412&     35.67&      2.05&      4.23&      0.58& 4904&  222& 1.30& 0.39&
 2.47&  0.03&  9.77&   0.52 &  0.990&  7.02&   0.00&   0.00 \\
2860936&    102.01&      5.63&      7.59&      0.84& 5062&  144& 3.07& 0.75&
 2.93&  0.03&114.32&   5.55 &  1.000&  3.91&   0.00&   0.00 \\
2988655&     30.63&      1.70&      3.87&      0.61& 5158&  313& 1.27& 0.43&
 2.42&  0.03& 71.85&   6.93 &  0.990&  7.67&   0.00&   0.00 \\
3102990&     29.77&      1.86&      3.79&      0.56& 4835&  177& 1.15& 0.36&
 2.39&  0.03& 85.44&   7.84 &  1.000&  7.84&   0.00&   0.00 \\
3216467&     31.47&      1.96&      3.83&      0.12& 4623&  154& 1.21& 0.16&
 2.40&  0.03&151.87&   8.36 &  0.990&  7.75&   0.00&   0.00 \\
3220837&    132.56&     20.10&     13.21&      0.31& 5294&  260& 0.79& 0.21&
 3.06&  0.07& 79.73&   7.54 &  0.990&  2.25&   0.00&   0.00 \\
3240280&     29.19&      1.71&      3.78&      1.71& 4810&  121& 1.08& 0.99&
 2.38&  0.03&103.83&   5.30 &  0.990&  7.86&   0.00&   0.00 \\
3324186&     32.44&      2.20&      3.91&      0.09& 4724&  110& 1.26& 0.16&
 2.42&  0.03& 51.87&   4.32 &  0.980&  7.60&   0.00&   0.00 \\
3432732&     80.48&      3.81&      6.45&      0.13& 5015&  150& 2.85& 0.28&
 2.83&  0.02&121.75&   5.62 &  1.000&  4.60&   0.00&   0.00 \\
3437031&     63.45&      2.74&      5.60&      0.18& 5059&  214& 2.49& 0.28&
 2.73&  0.02&138.77&   7.22 &  0.990&  5.30&   0.00&   0.00 \\
3439466&     29.33&      1.65&      3.91&      0.11& 4991&  235& 1.01& 0.13&
 2.39&  0.03& 46.75&   4.41 &  0.990&  7.60&   0.00&   0.00 \\
3448282&     30.45&      1.54&      3.59&      0.09& 4891&  241& 1.55& 0.18&
 2.40&  0.03& 55.98&   4.69 &  0.990&  8.27&   0.00&   0.00 \\
3526625&     34.35&      2.51&      4.18&      0.15& 4883&  184& 1.21& 0.18&
 2.45&  0.04&125.11&   5.95 &  1.000&  7.11&   0.00&   0.00 \\
3532985&     94.84&     10.80&      7.92&      0.27& 4752&  122& 1.89& 0.40&
 2.89&  0.05&  6.03&   0.29 &  0.990&  3.75&   0.00&   0.00 \\
3557606&    100.96&      6.24&      8.09&      0.24& 5044&  118& 2.29& 0.29&
 2.93&  0.03& 26.34&   2.99 &  0.990&  3.67&   0.00&   0.00 \\
3642135&     51.25&      2.53&      4.96&      0.15& 5060&  175& 2.13& 0.24&
 2.64&  0.02&144.72&   7.10 &  1.000&  5.99&   0.00&   0.00 \\
3750783&    110.02&      4.91&      8.68&      0.20& 5249&  300& 2.38& 0.27&
 2.97&  0.03&  6.79&   0.45 &  0.990&  3.42&   0.00&   0.00 \\
3758731&     74.83&      5.98&      6.28&      0.18& 5119&  248& 2.63& 0.42&
 2.80&  0.04&115.12&   3.69 &  1.000&  4.73&   0.00&   0.00 \\
3937217&     30.06&      1.94&      3.76&      0.11& 4887&  185& 1.24& 0.17&
 2.40&  0.03& 54.08&   4.43 &  1.000&  7.90&   9.08&   1.42 \\
3956210&     29.67&      3.21&      3.85&      0.50& 5015&  214& 1.13& 0.37&
 2.40&  0.05& 54.83&   4.94 &  0.990&  7.71&   0.00&   0.00 \\
4041075&     32.20&      2.00&      3.78&      0.13& 4290&  156& 1.23& 0.17&
 2.40&  0.03&111.25&   5.14 &  1.000&  7.86&   0.00&   0.00 \\
   ...&      \multicolumn{2}{c}{...}&      \multicolumn{2}{c}{...}&      \multicolumn{2}{c}{...}&      \multicolumn{2}{c}{...}&      \multicolumn{2}{c}{...}&      \multicolumn{2}{c}{...}&      ...&     ...\\
\hline
  \end{tabular}
   \begin{tablenotes}
  \item    {Note:} The crowding value represents the amount of the integrated flux belonging to the targeted star. Hence, it is equal to 1 when only the targeted star is in the aperture.
  \end{tablenotes}

\end{threeparttable}
\end{table*}

\begin{table*}
  \caption{\label{Tab:Sample2} Stars showing rotational modulation in their light curve, but probably due to  pollution from a nearby star.}
  \begin{threeparttable}
  \centering
  \begin{tabular}{c*{6}{r@{$\ \pm\ $}l}ccc}
KIC & \multicolumn{2}{c}{$\nu_\text{max}$ [$\mu$Hz]} &
\multicolumn{2}{c}{$\Delta\nu$ [$\mu$Hz]} &
\multicolumn{2}{c}{T$_{\rm{eff}}$ [K]} & 
\multicolumn{2}{c}{$\log g$ [dex]} & 
\multicolumn{2}{c}{M [M$_\odot$]} & 
\multicolumn{2}{c}{$P_\text{rot}$ [days]} & 
crowding & $T_\text{crit}$ & Flag \\
\hline
1164356&     27.80&      1.80&      3.33&      0.09& 5225&  169& 2.38& 0.03&
 1.76&  0.23& 15.21&   1.18 &  0.770&  8.92&  1 \\
1870433&    229.14&     25.83&     16.14&      0.52& 5099&  191& 3.29& 0.05&
 1.72&  0.36& 90.81&   4.40 &  0.970&  1.84&  1 \\
2157901&     28.89&      2.70&      3.93&      0.11& 4998&  277& 2.38& 0.05&
 0.95&  0.18& 43.92&   3.37 &  0.650&  7.56&  1 \\
2305407&    139.13&     20.11&     13.07&      0.30& 5017&  127& 3.07& 0.07&
 0.87&  0.22& 92.15&   4.31 &  0.980&  2.27&  1 \\
2436944&     30.40&      1.74&      3.75&      0.11& 4835&  192& 2.40& 0.03&
 1.27&  0.16&135.06&   5.78 &  0.860&  7.92&  1 \\
2437653&     74.10&      3.40&      7.03&      0.21& 4386&  130& 2.76& 0.02&
 1.29&  0.14& 10.97&   1.03 &  0.880&  4.22&  1 \\
2437987&     30.32&      1.80&      3.68&      0.10& 4708&  135& 2.39& 0.03&
 1.31&  0.16&113.56&   5.68 &  0.880&  8.07&  1 \\
2438051&     30.89&      1.64&      3.66&      0.11& 4724&  139& 2.40& 0.03&
 1.42&  0.16& 80.83&   7.30 &  0.970&  8.12&  1 \\
2570214&     25.44&      1.65&      3.65&      0.08& 4735&  148& 2.32& 0.03&
 0.81&  0.10& 69.40&   5.22 &  0.970&  8.14&  1 \\
2570518&     46.10&      2.70&      4.94&      0.15& 4399&  130& 2.56& 0.03&
 1.28&  0.16& 10.11&   0.73 &  0.870&  6.01&  1 \\
2719113&     32.58&      1.80&      3.96&      0.12& 4695&  122& 2.42& 0.03&
 1.21&  0.14& 94.80&   9.70 &  0.970&  7.50&  1 \\
2720444&     32.03&      2.72&      4.08&      0.10& 5013&  233& 2.43& 0.04&
 1.12&  0.19& 67.04&   7.32 &  0.980&  7.28&  1 \\
2833697&     30.95&      2.14&      4.11&      0.13& 5117&  126& 2.42& 0.03&
 1.01&  0.14& 42.13&   2.85 &  0.950&  7.23&  1 \\
3234655&     41.57&      2.36&      4.52&      0.13& 5233&  377& 2.55& 0.03&
 1.74&  0.25&128.09&   5.07 &  0.980&  6.57&  1 \\
3240573&     31.80&      2.28&      3.89&      0.12& 4752&  166& 2.41& 0.03&
 1.23&  0.18& 68.51&   3.52 &  0.970&  7.64&  1 \\
   ...&      \multicolumn{2}{c}{...}&      \multicolumn{2}{c}{...}&      \multicolumn{2}{c}{...}&      \multicolumn{2}{c}{...}&      \multicolumn{2}{c}{...}&      \multicolumn{2}{c}{...}&      ...&     ...\\
\hline
  \end{tabular}
  \begin{tablenotes}
  \item    {Note:} The value of the flag is equal to 1 if the star has been discarded
    because of a low crowding value, and 2 if it has been discarded
    because the ratio $P_\text{rot}/T_\text{crit}$ is too small  (see
    Section~\ref{Sec:surf} for details).
  \end{tablenotes}
\end{threeparttable}
\end{table*}

\subsection{Comparison with $v \sin(i)$ measurements}

The most common alternate method for studying stellar surface rotation is measuring the rotational line broadening $v\sin(i)$ through spectroscopic
observations. Such measurements are completely independent from our
analysis and allow us to verify whether the rotation periods we extract are
compatible with other observations. Moreover, we can estimate the
radii of the stars of our sample from their seismic global parameters
-- $\nu_{max}$ and $\Delta \nu$ -- and their effective temperatures
$T_\mathrm{eff}$ -- taken from \citet{2014ApJS..211....2H}. This then allows
the estimation of the inclination angle $i$, a parameter that is
otherwise difficult to constrain except via detailed seismic analysis
or  modelling  the transits for multiple systems.

\begin{figure}
\begin{center}
\subfigure{\includegraphics[width=9cm,clip ]{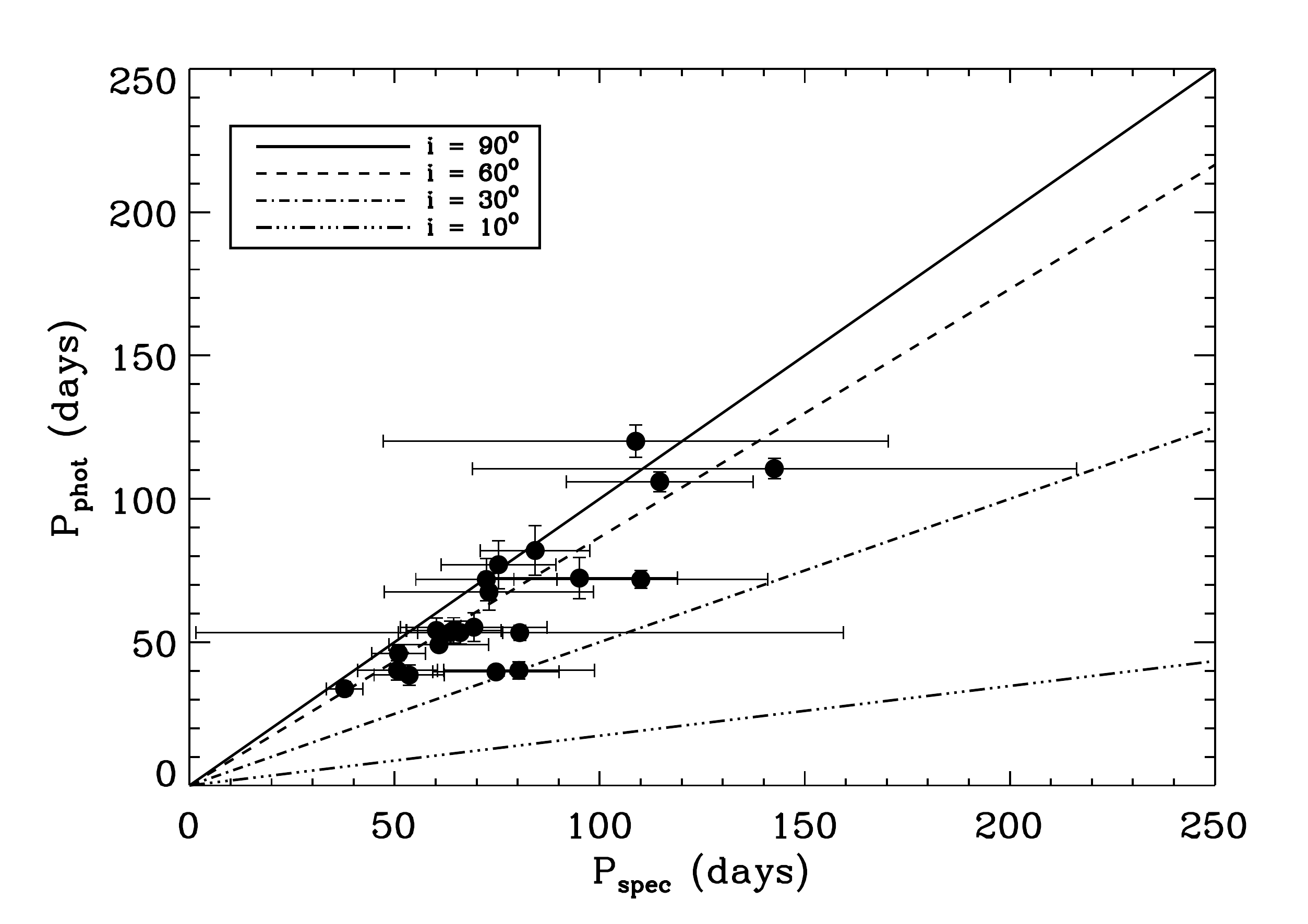}}
\subfigure{\includegraphics[width=9cm,clip]{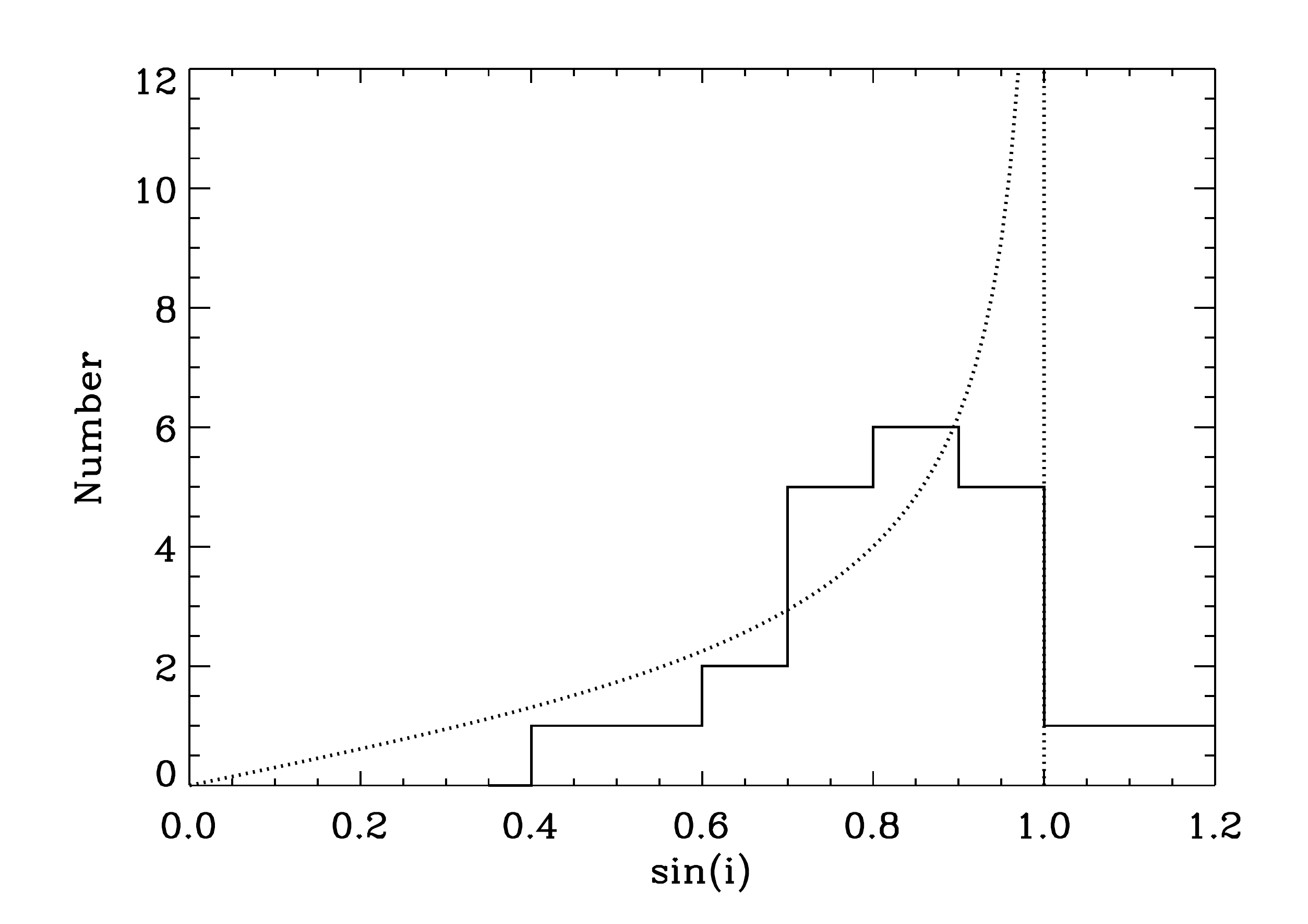}}
\caption{Top: Comparison between the $P_{rot}$ from this work and the
maximum period $P_{spec}$ compatible with $v sin(i)$ measured by
\citet{Tayar2015}. The different lines indicate where a star is
expected to fall for a given inclination angle $i$. Bottom:
Distribution of the $sin(i)$ for the rotation sample. The dotted line
corresponds to the distribution expected for a random distribution of inclinations.}
\label{Fig:vsini}
\end{center}
\end{figure}

Figure~\ref{Fig:vsini} presents the comparison between the $P_\text{rot}$
from this work with the maximum period possible for the $v\sin(i)$
derived by \citet{Tayar2015} from APOGEE spectra. It shows that
the periods we derive are fully compatible with the spectroscopic
measurements. Moreover, the distribution of $\sin(i)$ is relatively
compatible with a uniform random distribution of the angles $i$. This suggests that our sample of rotating red giants is not biased towards particular inclination angles.

\subsection{Binarity}

   {While isolated giants are not generally expected to have measurable spots, the same is not true for giants in close, tidally interacting binary systems \citep[e.g.][]{2014ApJ...785....5G,Beck2017b}. Similar to the W Ursa Majoris stars, tidal interactions tend to enhance surface activity and increase rotation rates.}
While not all of the stars in our sample have multiple epochs of spectroscopic observations, 
we searched the 116 stars in our validated sample with multiple APOGEE spectra for radial velocity variability greater than 1~km s$^{-1}$. That threshold is  larger than both the detection limit for this instrument
(0.5~km/s; \citealt{Deshpande2013}) and the expected radial velocity
jitter for red
giants (a surface gravity dependent quantity which can be as high as
0.5~km/s at a $\log(\text{g})$ of 1; \citealt{Hekker2008}).
We find significant radial
velocity variability (greater than 1~km/s) in six stars (5.2\% of the
searchable sample; KIC~5382824, 5439339, 6032639, 6933666, 7531136, and 12314910); two others (KIC
7661609 and 9240941) have suggestive variations (greater than
0.5~km/s, less than 1 km/s). Although very unlikely, it could be possible that the periods measured
for these stars are actually the rotation periods of their lower mass
companions. However, we suspect that in most cases the secondary is much smaller and therefore unlikely to substantially contribute to the variability of the blended source. Moreover, the rotation rates found for these giants are very similar to those found for the rest of the sample, which suggests that they are due to spots on the primary.
\subsection{Causes of rapid rotation in single stars}


  Because the stars in our sample have measured masses, we want to compare the distribution of active stars we measure to previous measurements of rapid rotation and to population synthesis models to determine whether our rapidly rotating single stars are massive stars born with rapid rotation or stars that  gained angular momentum through an interaction. Spectroscopic surveys indicate that about 2\%\ of red giants are rotating rapidly \citep[vsini$>$ 10 km s$^{-1}$,][]{2011ApJ...732...39C}. While it is difficult to directly compare our period detection fraction to these spectroscopic
rotation predictions because of unknown factors like the interplay
between rotation and magnetic excitation, we suspect that given
the observed correlations between rotation and activity \citep[e.g.][]{Noyes1984, MamajekHillenbrand2008} the
fraction of stars with detected photometric periods should be similar
to the fraction of rapidly rotating stars. Indeed, we find that we
detect periodic modulations in 2.08\% (361/17,377 ) of our sample, which is
consistent with    {the fraction of spectroscopically measured rapid rotators.} We therefore assume for the following analysis that the fraction of active stars we measure is related to the fraction of rapid rotators measured by spectroscopic methods. We note that our actual sample is the fraction of stars active enough to measure rotation periods and that we have no information on the distribution of the rotation periods in inactive stars.

\begin{figure}[!htb]
\begin{center}
\subfigure{\includegraphics[width=9cm, trim={1.5cm 0.5cm 0 8.8cm},clip]{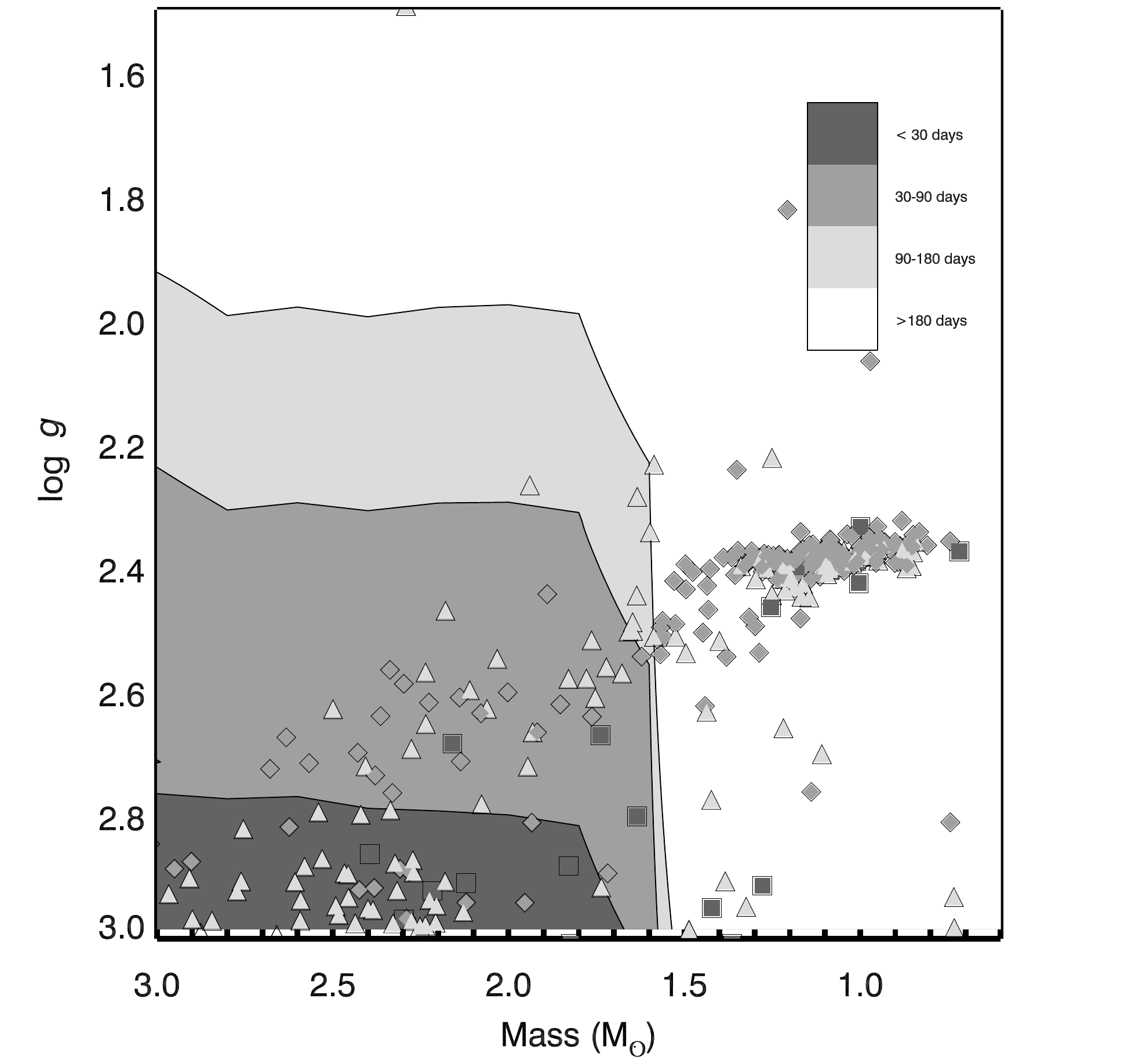}}
\subfigure{\includegraphics[width=9cm, trim={1.5cm 0.5cm 0 8.8cm},clip]{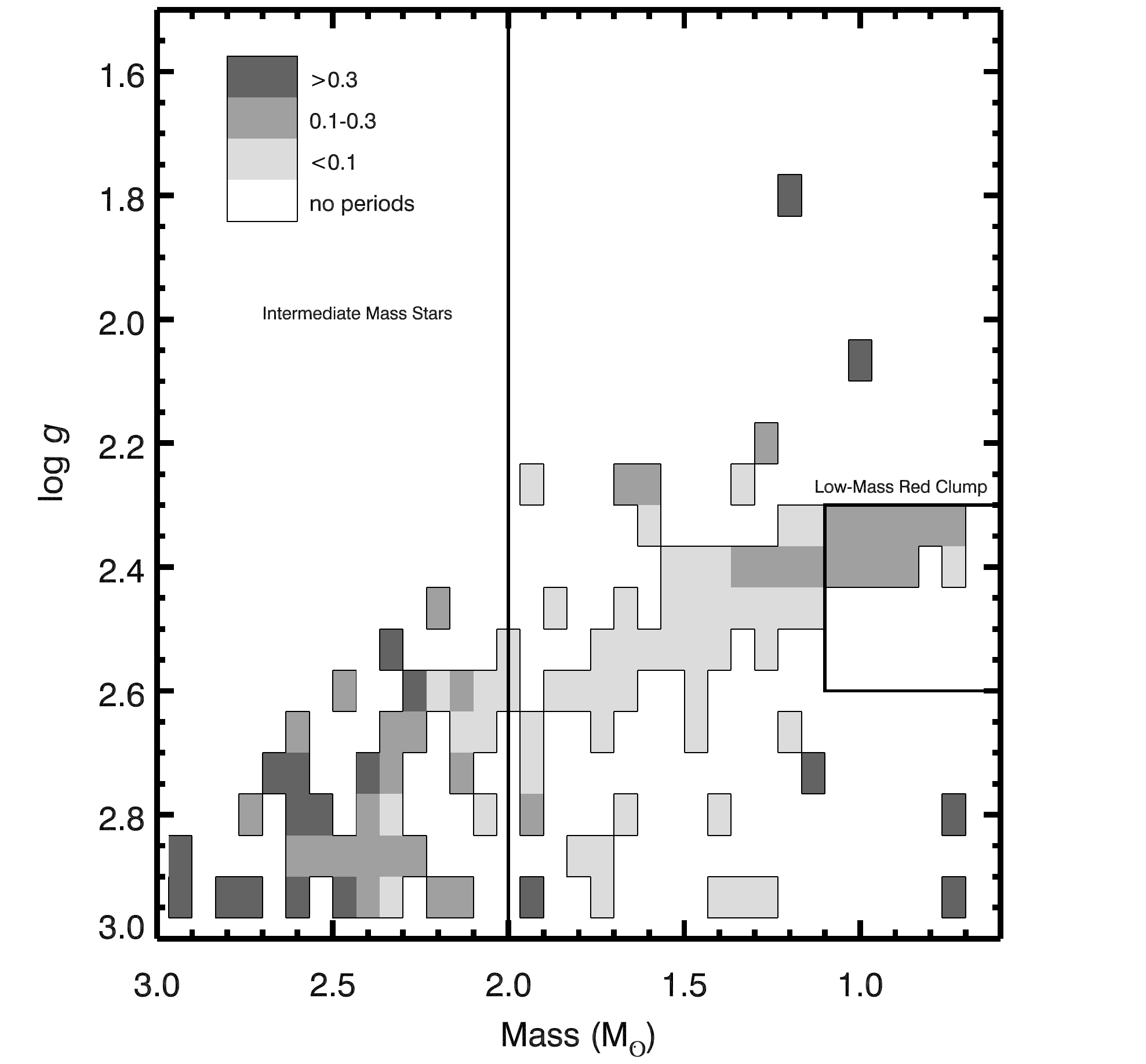}}
\caption{   {Top: Background colours indicating the prediction for the minimum rotation period expected for single stars from the models in  \citet{Tayar2015}:   $<$30 days in dark grey, 30-90 days in medium grey, 90-180 days in light grey, and $>$180 days in white. For comparison, we overplot the measured rotation periods for stars in our sample using the same period bins as dark grey squares, medium grey diamonds, and light grey triangles. Bottom: Fraction of stars with detected rotation periods in bins of mass and surface gravity. Dark grey indicates bins where $ >30\%$ of the stars had detected rotation, medium grey indicates bins with $10-30\%$ detections, light grey indicates bins where fewer than $10\%$ of stars had detected rotation periods, and white indicates bins with either no stars or no detected rotation periods. We have also indicated the regions we define as `intermediate mass' and `low-mass red clump'.}}
\label{Fig:sampleab}
\end{center}
\end{figure}

   Population synthesis models indicate that one to two percent of stars should be rapidly rotating from recent interactions or mergers with a companion star \citep{2011ApJ...732...39C}. Additionally, angular momentum conservation in intermediate mass stars predicts that a small percent of additional field giants should be rotating rapidly because they have not yet spun down. Because our stars have measured masses and surface gravities, we focus  on two regions of parameter space (see Figure \ref{Fig:sampleab}).

The first region is the low-mass red clump. These are stars that rotated slowly on the main sequence and therefore (see Figure \ref{Fig:sampleab}, top) they cannot have rotation detected at the periods we investigate on the giant branch unless they gain angular momentum from an interaction with another object.   It has recently been suggested that more than 7\%
of low-mass red clump stars are spun up as the result of an
interaction \citep{Tayar2015}. We detect rotation in 88 (15.3\%) of the 575 \emph{Kepler} giants with masses below $1.1 M_\odot$ and surface gravities between 2.3 and 2.6 dex, which is twice this previous
measurement. We therefore suggest that a large fraction of red giants have
undergone an interaction on the giant branch that spins up their surface and that the unexpectedly low number of fast rotators is due  to an overestimation of the stellar merger rates. Additionally, although it is beyond the scope of this paper, it would be
very interesting to study these red giants in more detail and especially to measure their surface abundances to check whether there is an enrichment in certain elements due to this interaction.

The second region  we analysed contains the intermediate mass stars. Observations indicate that intermediate mass stars
($M>2.0\ M_\odot$) have a wide range of rotation rates on the main
sequence \citep{Zorec2012}.    Assuming solid body rotation and solar-like angular momentum loss, more than 50\% of intermediate mass stars should still have rotation velocities above 10 km/s on the giant branch.  In contrast with such predictions, we find a smaller rate of rotating stars above two solar masses (94 out of 4881    {\emph{Kepler} giants in this mass range} or 1.92\%), and find that rotation is detected only in the smallest stars in this mass range (see Figure \ref{Fig:sampleab}, bottom).    This suggests that intermediate mass stars are either losing more angular momentum than a standard Kawaler wind loss law would predict \citep{1988ApJ...333..236K}, or they are undergoing a substantial amount of radial differential rotation. Seismic measurements of the core rotation of a few such stars indicate that radial differential rotation is occurring \citep[e.g.][]{2012Natur.481...55B,Deheuvels2015}, but more work should be done to characterize the extent of the differential rotation and the extent to which additional loss is required. We therefore suggest that the low fraction of active stars in our sample is due to an incorrect estimation of the fraction of rapidly rotating intermediate mass stars.

\section{Conclusions}
\label{Sec:Conclu}

We study a sample of 17,377 red giants with measured solar-like
oscillations from the \emph{Kepler} observations. We use various
techniques to detect the active stars in this sample and measure
their surface rotation rates from modulations of their light
curves. After carefully taking into account possible pollutants,
we extract a subsample of 361 red giants with accurate surface
rotation periods.

These red giants are peculiar in the sense that they show high
activity and rapid rotation. While we assume in this analysis that activity and rotation are correlated, we re-emphasize that we do not measure the distribution of rotation periods in inactive stars. However, we suspect that most of the inactive stars are rotating slowly as a result of the expansion of their outer layers and the extraction
of angular momentum through magnetized winds during the main sequence. 
We therefore assert that the majority of the active rapidly rotating stars in our sample
must have undergone an event that led to an acceleration of
their surface rotation.

Our detection rate of 2.08\% is indeed in very good agreement with
binary interactions predictions from \citet{Carlberg2011}. Moreover,
this rate increases to 15.3\% if we consider only the low-mass red
clump stars in our sample, which shows that red giants that have gone
through the whole red giant branch have a higher probability of having
undergone an interaction with a companion star or planet, as
suggested by \citet{Tayar2015}.

   However, when we consider only more massive stars that did not lose angular momentum on the main sequence and should therefore be rotating rapidly, we do not see the enhanced rate of detections that we  expected. This suggests that the discrepancy between the predicted and measured rates of rapid rotation in the field comes from the overestimation of the surface rotation rates of intermediate mass red giants. It indicates a complexity to the angular momentum transport and loss in these stars that is not currently taken into account and we suggest that more work should be done to understand how these stars differ from solar-type dwarfs.

This work opens the path to a large number of studies about red giant
stars. It can help to better understand the links between
activity and rotation for these objects as it offers a sample of
active and rapidly rotating red giants to the stellar community. In
particular, it would be interesting to see if any evidence can be
obtained to show that the red clump stars from our subsample have indeed
undergone an interaction with a companion.

\begin{acknowledgements} 
The authors wish to thank the entire \emph{Kepler} team, without whom
these results would not be possible. Funding for this Discovery
mission is provided by NASA's Science Mission Directorate. The authors
also wish to thank R.-M.~Ouazzani for very fruitful
discussions on this work. The authors acknowledges the KITP staff of
UCSB for their hospitality during the research program Galactic
Archaeology and Precision Stellar Astrophysics. M.H.P. and J.T. acknowledge
support from NSF grant AST-1411685. T.C., D.S., and R.A.G. received funding
from the CNES GOLF and PLATO grants at CEA. R.A.G. and P.G.B. acknowledge the
ANR (Agence Nationale de la Recherche, France) program IDEE
(ANR-12-BS05-0008) ``Interaction Des \'Etoiles et des
Exoplan\`etes''. S.M. acknowledges support from NASA grants NNX12AE17G 
and NNX15AF13G and NSF grant AST-1411685. The research
leading to these results has received funding from the European
Communitys Seventh Framework Programme ([FP7/2007-2013]) under grant
agreement no. 269194 (IRSES/ASK).
\end{acknowledgements} 

\bibliographystyle{aa}
\bibliography{./BIBLIO}
\end{document}